\documentstyle[12pt,subeq,axodraw]{article}

\newcommand\vek[1]{\mbox{\rmfamily\bfseries\itshape#1}}
\newcommand\vekexp[1]{\mbox{\scriptsize\rmfamily\bfseries\itshape#1}}
\def\rd{{\rm d}}

\begin{document}

\title{Thermodynamic Properties of the Two-Dimensional
Two-Component Plasma}
\author{L. \v Samaj \ and \ I. Trav\v enec \\ 
Institute of Physics, Slovak Academy of Sciences,\\
D\' ubravsk\' a cesta 9, 842 28 Bratislava, Slovakia;\\
e-mail: fyzimaes@savba.sk (L.\v{S}.), fyzitrav@savba.sk (I.T.)}

\maketitle

\begin{abstract}
The model under consideration is a two-dimensional two-component plasma,
stable against collapse for the dimensionless coupling constant
$\beta < 2$.
The combination of a technique of renormalized Mayer expansion with 
the mapping onto the sine-Gordon theory provides the full thermodynamics
of the plasma in the whole stability range of $\beta$.
The explicit forms of the density-fugacity relationship and of 
the specific heat (at constant volume) per particle are presented.
\end{abstract}

\vfill


\newpage

\section{Introduction}
The model under consideration is the two-dimensional (2D) two-component
plasma (TCP), i.e., a neutral system of pointlike particles of positive
and negative charge, interacting through the 2D (logarithmic) Coulomb
interaction.
Classical equilibrium statistical mechanics is applied.
The system is stable against collapse for the dimensionless coupling
constant $\beta$ (sometimes denoted as $\Gamma$) lower than 2.
The absence of some short-range interaction, e.g. the hard-core one,
prevents to extend the treatment beyond $\beta =2$ and to reach
the Kosterlitz-Thouless phase transition which occurs around $\beta=4$.

In arbitrary dimension, the long-range tail of the Coulomb potential
causes screening, and thus gives rise to exact constraints
for the charge-charge structure function (see review \cite{Martin}).
In the bulk regime, the zeroth- and second-moment conditions for the
structure function are known exactly \cite{Stillinger1}, \cite{Stillinger2}.
The specialization to the 2D TCP brings additional sum rules.
The knowledge of the exact equation of state \cite{Hauge} fixes via
the compressibility equation the zeroth moment of the {\it number density}
correlation function \cite{Vieillefosse}.
In a recent paper \cite{Jancovici1}, the second-moment formula for that
density-density correlation function has been derived by using analogies
with critical systems.
This result was rederived directly by using a renormalized Mayer expansion
\cite{Deutsch}, \cite{Kalinay} in Ref. \cite{Jancovici2}.

At arbitrary temperature, the 2D TCP is supposed to be in the critical
state in terms of the induced electrical-field correlations (but not
the particle correlations) \cite{Forrester}, \cite{Jancovici3}.
The free energy therefore exhibits a universal finite-size correction
predicted by the Conformal Invariance theory.

The 2D TCP is mappable onto the Quantum Field Theory (QFT) models,
namely the 2D Thirring model (see e.g. \cite{ZinnJustin}) and the 2D
Euclidean, or (1+1)-dimensional quantum, sine-Gordon model \cite{Edwards}, 
\cite{Brydges}, \cite{Minnhagen}.
The integrability of the sine-Gordon theory was established and
analysed in Refs. \cite{Faddeev}, \cite{AZamolodchikov}, \cite{Destri},
\cite{AlZamolodchikov}, \cite{Lukyanov} and in many others.

This paper is devoted to the derivation of the thermodynamics
of the 2D TCP in the whole stability interval $0\le \beta <2$.
The aim is reached via a study of the relationship between 
the fugacity $z$ and the total number density of particles $n$.
It is known from a scaling argument that the combination
$n^{(1-\beta/4)}/z$ is a function of $\beta$ only.
Up to now, the exact form of the $n-z$ relationship was known only in
the Debye-H\"uckel $\beta \to 0$ limit, $n/z \to 2$,
and at the exactly solvable collapse border $\beta = 2$ \cite{Gaudin},
\cite{Cornu}, when $\sqrt{n}/z \to \infty$.
The little interest in the topic in the past is surprising in view of the
fact that in the QF theories related to the 2D TCP one-point expectations
of local fields are the objects of primary importance.
In the Conformal Perturbation treatment of a QFT, all multiple correlation
functions are reducible by the Operator-Product-Expansion \cite{Wilson}
to the one-point vacuum expectations which contain all the non-perturbative
information about the QFT.

We first investigate the $n-z$ relationship by using the renormalized
Mayer expansion for many-component fluids \cite{Jancovici2} and find out
that $n^{(1-\beta/4)}/z \sim \beta^{\beta/4} \times$ an analytic
function of solely $\beta$, $f(\beta)$, whose Taylor expansion around
$\beta = 0$ can be constructed systematically by evaluating certain
convergent integrals of Bessel functions.
The mapping onto the 2D sine-Gordon theory, whose normalization of the
relevant $\cos$-field is consistent with the formalism of the
renormalized Mayer expansion, then provides the explicit form of the
function $f(\beta)$, checked on a few lower orders of its $\beta$-expansion.
The exact density-fugacity relationship provides the full thermodynamics 
of the 2D TCP. 

The paper is organized as follows.
Section 2 recapitulates briefly the ordinary Mayer formalism for 
many-component fluids. 
The procedure of the renormalization of the Mayer expansion is described
in Section 3.
The application to the 2D TCP is presented in Section 4.
The information gained from the mapping onto the 2D sine-Gordon QFT
is the subject of Section 5.
The explicit form of the specific heat (at constant volume) per particle 
is given together with a discussion of the results in Section 6. 

\section{Ordinary Mayer expansion for many-component fluids}
We first consider a general classical multi-component fluid in
thermodynamic equilibrium at temperature $T$.
The last system is composed of distinct species of particles $\{ \sigma \}$.
The particles may be exposed to an external potential 
$u({\vek r},\sigma)$ and interact with each other through the pair
potential $v({\vek r}_i,\sigma_i\vert {\vek r}_j,\sigma_j)$ which depends
on the mutual distance $\vert {\vek r}_i - {\vek r}_j \vert$ of particles
$i, j$ as well as on their types $\sigma_i, \sigma_j$.
Vector position ${\vek r}_i$ of a particle in $d$-dimensional space
will be sometimes represented simply by $i$.
In the grand canonical formalism, denoting by $\{ \mu_{\sigma} \}$
the chemical potentials of species $\{ \sigma \}$, the grand-canonical
partition function is defined by
\begin{equation} \label{1}
\Xi(T,V,\{ \mu_{\sigma} \}) = \sum_{\{ N_{\sigma} \}} \prod_{\sigma}
{1\over N_{\sigma}!} \int_V \prod_{i=1}^N
\left[ \rd i ~ z(i,\sigma_i) \right]
\exp\left[ -{\beta \over 2} \sum_{(i\ne j)=1}^N v(i,\sigma_i \vert j,\sigma_j)
\right]
\end{equation}
where the first sum runs over all possible species numbers,
$N=\sum_{\sigma}N_{\sigma}$ is the total particle number,
$\beta = 1/kT$ the inverse temperature and
$z({\vek r},\sigma) = \exp [\beta\mu_{\sigma} - \beta u({\vek r},\sigma)]$
denotes the fugacity of $\sigma$-particles.
In the direct format, i.e. with fugacities $\{ z({\vek r},\sigma) \}$
as controlling variables, $\Xi$ is the generator for the particle
densities $\{ n({\vek r},\sigma) \}$ in the sense that
\begin{equation} \label{2}
n({\vek r},\sigma) = z({\vek r},\sigma) {\delta \ln \Xi[z] \over
\delta z({\vek r},\sigma)}
\end{equation}

The transition to the inverse format, i.e. with densities
$\{ n({\vek r},\sigma) \}$ as controlling variables, is based on
the Legendre transformation
\begin{equation} \label{3}
-\beta {\bar F}[n] = \ln \Xi - \int_V \rd {\vek r} \sum_{\sigma}
n({\vek r},\sigma) \ln z({\vek r},\sigma)
\end{equation}
which defines the Helmholtz free energy ${\bar F}$ as the explicit
density functional.
The subtraction of the one-particle part provides the dimensionless
(for notational convenience, minus) excess free energy
\begin{equation} \label{4}
\Delta[n] = -\beta {\bar F}[n] + \int_V \rd {\vek r} \sum_{\sigma}
\left[ n({\vek r},\sigma) \ln n({\vek r},\sigma) - n({\vek r},\sigma)
\right]
\end{equation}
It is easy to show that $\Delta[n]$ is the generator in the following
sense
\begin{equation} \label{5}
\ln \left[ {n({\vek r},\sigma) \over z({\vek r},\sigma)} \right]
= {\delta \Delta[n] \over \delta n({\vek r},\sigma)}
\end{equation}
Its Mayer diagrammatic representation in density reads \cite{Hansen}
\begin{eqnarray} \label{6}
\Delta[n] & = & \big\{ {\rm all\ connected\ diagrams\ which\ consist\ of\ }
N\ge 2\ {\rm field\ } n(i,\sigma_i)-{\rm circles}  \nonumber \\
& & \quad {\rm and\ } f(i,\sigma_i \vert j,\sigma_j)-{\rm bonds,\ 
and\ are\ free\ of\ connecting\ circles} \big\} 
\end{eqnarray}
(the removal of a connecting circle disconnects the diagram).
Here,
\begin{equation} \label{7}
f(i,\sigma_i \vert j,\sigma_j) = \exp \left[ -\beta v(i,\sigma_i \vert
j,\sigma_j) \right] - 1
\end{equation}
is called the Mayer function and, besides the integration over spatial
coordinate of a field (black) circle, the summation over all 
$\sigma$-states at this vertex is assumed as well.

\section{Renormalized Mayer expansion}
The renormalized Mayer representation of $\Delta[n]$ results from
Eq.(\ref{6}) in two steps \cite{Deutsch}, \cite{Kalinay}:

\noindent i) the expansion of each Mayer function in the inverse temperature,
\begin{equation} \label{8}
f(1,\sigma_1\vert 2,\sigma_2) = - \beta v(1,\sigma_1\vert 2,\sigma_2)
+ {1\over 2!} \left[ - \beta v(1,\sigma_1\vert 2,\sigma_2) \right]^2
+ \ldots
\end{equation}
or, graphically,
\vspace{5pt} \hfill\\

\hskip 3truecm
\begin{picture}(55,40)(0,7)
    \Line(0,10)(50,10)
    \BCirc(0,10){2.5} \BCirc(50,10){2.5}
    \Text(0,0)[]{1,$\sigma_1$} \Text(50,0)[]{2,$\sigma_2$}
    \Text(25,23)[]{$f$}
\end{picture} $\ =\ \ $
\begin{picture}(55,20)(0,7)
    \DashLine(0,10)(50,10){7}
    \BCirc(0,10){2.5} \BCirc(50,10){2.5}
    \Text(0,0)[]{1,$\sigma_1$} \Text(50,0)[]{2,$\sigma_2$}
    \Text(24,23)[]{$-\beta v$}
\end{picture} $\ +\ \ $
\begin{picture}(55,25)(0,7)
    \DashCArc(25,-14)(34,45,135){5}
    \DashCArc(25,34)(34,225,315){5}
    \BCirc(0,10){2.5} \BCirc(50,10){2.5}
    \Text(0,0)[]{1,$\sigma_1$} \Text(50,0)[]{2,$\sigma_2$}
\end{picture} $\ +\ \ldots$
\hfill {(8')}
\vskip 0.7cm
\noindent where the factor 1/(number of interaction lines)! is automatically
assumed;

\noindent ii) the consequent series elimination of two-coordinated field 
circles between every couple of three- or more-coordinated field circles
(hereinafter, by coordination of a circle we mean its bond-coordination,
i.e. the number of bonds meeting at this circle).
The renormalized $K$-bonds are given by
\begin{equation} \label{9}
{\begin{picture}(55,20)(0,7)
    \Photon(0,10)(55,10){1}{7}
    \BCirc(0,10){2.5} \BCirc(55,10){2.5}
    \Text(0,0)[]{$1,\sigma_1$} \Text(55,0)[]{$2,\sigma_2$}
    \Text(28,23)[]{$K$}
\end{picture}}
\ \ \ = \ \ \
{\begin{picture}(55,20)(0,7)
    \DashLine(0,10)(55,10){7}
    \BCirc(0,10){2.5} \BCirc(55,10){2.5}
    \Text(0,0)[]{$1,\sigma_1$} \Text(50,0)[]{$2,\sigma_2$}
\end{picture}}\ \ +\ \
{\begin{picture}(110,20)(0,7)
    \DashLine(0,10)(55,10){5}
    \DashLine(55,10)(110,10){5}
    \BCirc(0,10){2.5} \BCirc(110,10){2.5}
    \Vertex(55,10){2.2}
    \Text(0,0)[]{$1,\sigma_1$} \Text(110,0)[]{$2,\sigma_2$}
\end{picture}}\ \ + \ldots
\end{equation}
or, algebraically,
$$K(1,\sigma_1\vert 2,\sigma_2) =  [-\beta v(1,\sigma_1 \vert 2, \sigma_2)]
+ \sum_{\sigma_3} \int_V \rd 3 ~ [-\beta v(1,\sigma_1 \vert 3,\sigma_3)]~
n(3,\sigma_3)~ K(3,\sigma_3 \vert 2,\sigma_2) \eqno(9')$$

The procedure of bond-renormalization transforms the ordinary
Mayer representation (\ref{6}) of $\Delta$ into \cite{Jancovici2}
\begin{subequations} \label{10}
\begin{equation} \label{10a}
\Delta[n] = \ \ 
\begin{picture}(50,20)(0,7)
    \DashLine(0,10)(40,10){5}
    \Vertex(0,10){2.2} \Vertex(40,10){2.2}
\end{picture}
 + \ D_0[n] + \ \sum_{s=1}^{\infty}\ D_s[n] ,
\end{equation}
where 
\begin{eqnarray} \label{10b}
D_0 & = & \ \
\begin{picture}(40,20)(0,7)
    \DashCArc(20,-10)(28,45,135){5}
    \DashCArc(20,30)(28,225,315){5}
    \Vertex(0,10){2} \Vertex(40,10){2}
\end{picture} \ \ +\ \ 
\begin{picture}(40,20)(0,19)
    \DashLine(0,10)(40,10){5}
    \DashLine(0,10)(20,37){5}
    \DashLine(20,37)(40,10){5}
    \Vertex(0,10){2} \Vertex(40,10){2} \Vertex(20,37){2}
\end{picture} \ \ +\ \ 
\begin{picture}(30,20)(0,10)
    \DashLine(0,0)(30,0){5}
    \DashLine(0,0)(0,30){5}
    \DashLine(0,30)(30,30){5}
    \DashLine(30,0)(30,30){5}
    \Vertex(0,0){2} \Vertex(30,0){2} \Vertex(0,30){2} \Vertex(30,30){2}
\end{picture} \ \ +\ \ \ldots  \nonumber \\
& & \\
& = & \sum_{N=2}^{\infty} {1\over 2N} \sum_{\sigma_1 \ldots \sigma_N}
\int_V \prod_{i=1}^N \left[ \rd i ~ n(i,\sigma_i) \right] 
\left[ -\beta v(1,\sigma_1\vert 2,\sigma_2) \right] \nonumber \\
& & \quad \quad \left[ -\beta v(2,\sigma_2\vert 3,\sigma_3) \right] \ldots
\left[ -\beta v(N,\sigma_N\vert 1,\sigma_1) \right] \nonumber
\end{eqnarray}
is the sum of all unrenormalized ring diagrams (which cannot undertake
the renormalization procedure because of the absence of three- or
more-coordinated field points) and
\begin{eqnarray} \label{10c}
\sum_{s=1}^{\infty} D_s & = & \big\{
{\rm all\ connected\ diagrams\ which\ consist\ of\ } N\ge 2 {\rm
\ field} \nonumber \\
& & n(i,\sigma_i){\rm -circles\ of\ (bond)\ coordination\ }\ge 3 
{\rm \ and\ multiple }  \nonumber \\
& &K(i,\sigma_i\vert j,\sigma_j){\rm -bonds,\ and\ are\ free\ 
of\ connecting\ circles\  \big\}}
\end{eqnarray}
\end{subequations}
represents the set of all remaining completely renormalized graphs.
By multiple $K$-bonds one means the possibility of an arbitrary
number of $K$-bonds between a couple of field circles, with the obvious
topological factor $1/({\rm number\ of\ bonds})!$. 
The order of $s$-enumeration is irrelevant, let us say
\begin{equation} \label{11}
\begin{picture}(60,40)(0,7)
    \PhotonArc(20,6)(20,15,165){1}{11}
    \PhotonArc(20,14)(20,195,345){1}{11}
    \Photon(0,10)(40,10){1}{8.5}
    \Vertex(0,10){2} \Vertex(40,10){2}
    \Text(20,-25)[]{$D_1$}
\end{picture}
\begin{picture}(60,40)(0,7)
    \PhotonArc(20,-10)(28,45,135){1}{9}
    \PhotonArc(20,30)(28,225,315){1}{9}
    \PhotonArc(20,6)(20,15,165){1}{11}
    \PhotonArc(20,14)(20,195,345){1}{11}
    \Vertex(0,10){2} \Vertex(40,10){2}
    \Text(20,-25)[]{$D_2$}
\end{picture}
\begin{picture}(60,40)(0,16)
    \PhotonArc(32,6)(32,115,170){1}{7.5}
    \PhotonArc(-12,40)(32,295,355){1}{7.5}
    \PhotonArc(9,5)(32,10,70){1}{7.5}
    \PhotonArc(52,40)(32,185,250){1}{7.5}
    \Photon(0,10)(40,10){1}{8}
    \Vertex(0,10){2} \Vertex(40,10){2} \Vertex(20,35){2}
    \Text(20,-15)[]{$D_3$}
\end{picture}
\SetScale{0.9}
\begin{picture}(60,40)(0,13)
    \Photon(0,0)(0,40){1}{7}
    \Photon(40,0)(40,40){1}{7}
    \Vertex(0,0){2} \Vertex(40,0){2} \Vertex(0,40){2} \Vertex(40,40){2}
    \PhotonArc(20,-20)(28,45,135){1}{8}
    \PhotonArc(20,20)(28,225,315){1}{8}
    \PhotonArc(20,20)(28,45,135){1}{8}
    \PhotonArc(20,60)(28,225,315){1}{8}
    \Text(20,-19)[]{$D_4$}
\end{picture} 
\begin{picture}(60,40)(0,12)
    \Photon(0,0)(40,0){1}{7}
    \Photon(0,0)(0,40){1}{7}
    \Photon(0,40)(40,40){1}{7}
    \Photon(40,0)(40,40){1}{7}
    \Photon(0,0)(40,40){1}{8}
    \Photon(0,40)(40,0){1}{8}
    \Vertex(0,0){2} \Vertex(40,0){2} \Vertex(0,40){2} \Vertex(40,40){2}
    \Text(20,-19)[]{$D_5$}
\end{picture}
\SetScale{1}
\begin{picture}(60,40)(0,16)
    \PhotonArc(32,6)(32,115,170){1}{7.5}
    \PhotonArc(-12,40)(32,295,355){1}{7.5}
    \PhotonArc(9,5)(32,10,70){1}{7.5}
    \PhotonArc(52,40)(32,185,250){1}{7.5}
    \PhotonArc(20,-17)(32,55,125){1}{7.5}
    \PhotonArc(20,39)(34.5,235,310){1}{9}
    \Vertex(0,10){2} \Vertex(40,10){2} \Vertex(20,35){2}
    \Text(20,-15)[]{$D_6$}
\end{picture}
\end{equation}
\vskip0.5cm
\noindent etc.

In accordance with relation (\ref{5}), $\ln [n(1,\sigma_1)/z(1,\sigma_1)]$ 
is expressible in the renormalized format as follows
\begin{subequations} \label{12}
\begin{equation} \label{12a}
\ln \left[ {n(1,\sigma_1) \over z(1,\sigma_1)} \right] = \quad \ 
\begin{picture}(50,20)(0,7)
    \DashLine(0,10)(40,10){5}
    \BCirc(0,10){2.2} \Vertex(40,10){2.2}
    \Text(-3,-2)[]{$1,\sigma_1$} 
\end{picture}
 + \ d_0(1,\sigma_1) + \ \sum_{s=1}^{\infty}\ d_s(1,\sigma_1)
\end{equation}
where $d_0(1,\sigma_1) = \delta D_0 / \delta n(1,\sigma_1)$ can be
readily obtained in the form
\begin{equation} \label{12b}
d_0(1,\sigma_1)  =  {1\over 2} \lim_{2\to 1} 
\left[ K(1,\sigma_1\vert 2,\sigma_2) + 
\beta v(1,\sigma_1\vert 2,\sigma_2) \right] \big\vert_{\sigma_2 = \sigma_1} 
\end{equation}
and
\begin{equation} \label{12c}
d_s(1,\sigma_1) = {\delta D_s \over \delta n(1,\sigma_1)}
\end{equation}
\end{subequations}
$(s=1,2,\ldots)$ denotes the whole family of $(1,\sigma_1)$-rooted
diagrams generated from $D_s$.
To get a family, one has to consider the functional dependence
of the dressed $K$-bonds (\ref{9}) on the species densities as well.
Since relation (9') yields
\begin{equation} \label{13}
{\delta K(1,\sigma_1\vert 2,\sigma_2) \over \delta n(3,\sigma_3)}
= K(1,\sigma_1\vert 3,\sigma_3) K(3,\sigma_3\vert 2,\sigma_2)
\end{equation}
the root (white) circle is generated, besides the field-circle positions,
also on $K$-bonds, causing their correct $K-K$ division.
For example, in the case of generator $D_1$ drawn in (\ref{11}),
one gets
\begin{equation} \label{14}
d_1(1,\sigma_1) =\ \ \ \ \
\begin{picture}(55,40)(0,7)
    \PhotonArc(20,6)(20,15,165){1}{11}
    \PhotonArc(20,14)(20,195,345){1}{11}
    \Photon(0,10)(40,10){1}{8.5}
    \BCirc(0,10){2.5} \Vertex(40,10){2.5}
    \Text(-7,-2)[]{$1,\sigma_1$} 
\end{picture} +\ \ \ \ \
\begin{picture}(55,40)(0,7)
    \PhotonArc(20,6)(20,15,165){1}{11}
    \PhotonArc(20,14)(20,195,345){1}{11}
    \Photon(0,10)(40,10){1}{8.5}
    \Vertex(0,10){2.5} \BCirc(20,26){2.5}
    \Text(20,36)[]{$1,\sigma_1$}
    \Vertex(40,10){2.2}
\end{picture}
\end{equation}

\section{The 2D TCP}
Let us now concentrate on the neutral 2D TCP of positive $(\sigma=+)$ and
negative $(\sigma=-)$ pointlike unit ($q=1$) charges, 
with the Coulomb interaction energy given by
\begin{subequations}
\begin{eqnarray} \label{15}
-\beta v(i,\sigma_i\vert j,\sigma_j) & = & \sigma_i \sigma_j 
\left[ -\beta v(i,j) \right] \\
v(i,j) & = & - \ln (\vert i-j \vert /L) 
\end{eqnarray}
\end{subequations}
where the constant $L$ is for simplicity set to unity and $\beta$ now
plays the role of the dimensionless coupling constant.
Considering the regime with homogeneous densities 
$n({\vek r},\sigma) = n_{\sigma}$,
the requirement of the charge neutrality implies $n_+ = n_- = n/2$
($n$ is the total number density of particles).
In the fugacity language, the chemical potentials of $\pm$ charged
particles must equal to one another, $z({\vek r},\sigma) = z({\vek r})$;
in the infinite-volume limit, $z({\vek r})$ acquires its bulk value
$z$.

The renormalized bonds, defined by (9'), now take the form
\begin{subequations} \label{16}
\begin{equation} \label{16a}
K(i,\sigma_i\vert j,\sigma_j) = \sigma_i \sigma_j K(i,j)
\end{equation}
where $K(i,j)$ satisfy the relation
\begin{equation} \label{16b}
K(1,2) = \left[ -\beta v(1,2) \right] + \int_V \rd 3 \left[ -\beta v(1,3)
\right] n ~ K(3,2)
\end{equation}
\end{subequations}
In the $V\to\infty$ limit, characterized by translationally 
invariant $K(i,j) = K(\vert i-j \vert )$, the Fourier transformation
of (\ref{16b}) results in
\begin{equation} \label{17}
{\hat K}(k) = \left[ -\beta {\hat v}(k) \right]
+ 2\pi n \left[ -\beta {\hat v}(k) \right] {\hat K}(k)
\end{equation}
Since the Fourier component ${\hat v}(k) = 1/k^2$, one arrives at
\begin{eqnarray} \label{18}
K(r) & = & - \beta \int {\rd^2 k \over 2\pi} {1\over k^2 + 
2\pi \beta n} \exp({\rm i}{\vek k} \cdot {\vek r}) \nonumber \\
& = & - \beta K_0(r\sqrt{2\pi\beta n})
\end{eqnarray} 
where $K_0$ is the modified Bessel function of second kind.
The special scaling form of $K$ has a fundamental impact on the $n$-
and $\beta$-classification of renormalized diagrams.

Let us first study the renormalized representation (\ref{10}) of 
the generator $\Delta(n)$:

\noindent (i) The first term on the rhs of (\ref{10a})
\begin{equation} \label{19}
{1\over 2!} \sum_{\sigma_1,\sigma_2=\pm 1} \int_V \rd 1 \rd 2 ~
n(1,\sigma_1) \sigma_1 \sigma_2 \left[ -\beta v(1,2) \right]
n(2,\sigma_2)
\end{equation}
is fixed to zero by the charge neutrality.

\noindent (ii) The second term $D_0$ (\ref{10b}) is expressible as follows
\begin{equation} \label{20}
D_0(n) = \sum_{N=2}^{\infty} {n^N\over 2N} \int_V \prod_{i=1}^N
\left[ -\beta v(1,2) \right] \left[ -\beta v(2,3) \right] 
\ldots \left[ -\beta v(N,1) \right]
\end{equation}
or, equivalently,
$$D_0(n) = {1\over 2} \int_0^n \rd n' \sum_{N=2}^{\infty} n'^{(N-1)} 
\int_V \prod_{i=1}^N \left[ -\beta v(1,2) \right] \left[ -\beta v(2,3) \right] 
\ldots \left[ -\beta v(N,1) \right] \eqno(20')$$
In the $V\to\infty$ limit, with respect to
(\ref{16b}), the sum over $N$
on the rhs of (20') is nothing but $V \times 
\lim_{r\to 0} [ K(r) + \beta v(r)]$ evaluated at $n=n'$.
Using the small-$x$ expansion of $K_0(x)$ \cite{Gradshteyn},
\begin{equation} \label{21}
K_0(x) = - \ln \left( {x\over 2} \right) I_0(x) +
\sum_{i=0}^{\infty} {x^{2i} \over 2^{2i} (i!)^2} \psi(i+1)
\end{equation}
where
$$I_0(x) = \sum_{i=0}^{\infty} {x^{2i} \over 2^{2i} (i!)^2}
\quad \quad {\rm and} \quad \quad
\psi(x) = {d\over dx} \ln \Gamma(x) $$
is the psi function $[\psi(1)=-C,\ C\ {\rm the\
Euler's\ constant}]$, one finally gets
\begin{equation} \label{22}
{D_0(n) \over V} = {\beta \over 4} (n \ln n - n) + {\beta n \over 2}
\left[ C + {1\over 2} \ln \left( {\pi \beta \over 2} \right) \right] 
\end{equation}

\noindent (iii) Let the given completely renormalized diagram 
$D_s$ $(s=1,2,\ldots)$, belonging to the sum (\ref{10c}), 
be composed of $N_s$ skeleton vertices
$i=1,\ldots,N_s$ of coordination $\nu_i \ge 3$ and $L_s$ bonds
$\alpha = 1,\ldots,L_s$.
$D_s$ can be formally expressed as
\begin{equation} \label{23}
D_s[n] = t_s \sum_{\sigma_1\ldots\sigma_{N_s}=\pm 1} \int_V
\prod_{i=1}^{N_s} \left[ \rd i ~ n(i,\sigma_i) \right] \prod_{\alpha=1}^{L_s}
K(\alpha_1,\sigma_{\alpha_1}\vert \alpha_2,\sigma_{\alpha_2})
\end{equation}
where $t_s$ is the topological factor and $\alpha_1, \alpha_2 \in
\{ 1,\ldots,N_s \}$, $\alpha_1 < \alpha_2$, denotes the ordered pair
of vertices joint by the $\alpha$-bond.
Since, according to (\ref{16a}),
$K(\alpha_1,\sigma_{\alpha_1}\vert \alpha_2,\sigma_{\alpha_2}) = 
\sigma_{\alpha_1} \sigma_{\alpha_2} K_{\alpha}$, $K_{\alpha} \equiv
K(\alpha_1,\alpha_2)$, Eq. (\ref{23}) can be rewritten as follows
\begin{equation} \label{24}
D_s[n] = t_s \sum_{\sigma_1\ldots\sigma_{N_s}=\pm 1} \int_V
\prod_{i=1}^{N_s} \left[ \rd i ~ n(i,\sigma_i) ~ \sigma_i^{\nu_i} \right] 
\prod_{\alpha=1}^{L_s} K_{\alpha}
\end{equation}
The set of coordination numbers $\{ \nu_i \}$ is constrained by
$\sum_{i=1}^{N_s} \nu_i = 2 L_s$ as every bond is shared by just two vertices.
For $n(i,\sigma_i) = n/2$, one has
\begin{equation} \label{25}
D_s(n) = t_s \left( {n\over 2} \right)^{N_s} 
\sum_{\sigma_1\ldots\sigma_{N_s}=\pm 1} \prod_{i=1}^{N_s} \sigma_i^{\nu_i}
\int_V \prod_{i=1}^{N_s} \rd i \prod_{\alpha=1}^{L_s} K_{\alpha}
\end{equation}
i.e., $D_s(n) \ne 0$ if and only if the coordinations of all vertices 
$\{ \nu_i \}$ in $D_s$ are even numbers $(\ge 4)$.
Let us suppose that this condition is fulfilled.
In the $V\to\infty$ limit, due to the invariance of the integrated
product $\prod_{\alpha=1}^{L_s} K_{\alpha}$ with respect to a uniform
shift in all integration variables $\{ i \}$, one of these variables
can be chosen as a reference put at the origin ${\vek 0}$, with the
simultaneous multiplication by volume $V$,
\begin{equation} \label{26}
D_s(n) = t_s n^{N_s} V \int \prod_{i=1}^{N_s} \rd i ~ \delta(j)
\prod_{\alpha=1}^{L_s} K_{\alpha}
\end{equation}
$j=1,\ldots,N_s$.
The scaling form of $K_{\alpha} = - \beta K_0(\vert \alpha \vert
\sqrt{2\pi\beta n})$ permits us to perform the $n$- and 
$\beta$-classification of the integral in (\ref{26}).
Every dressed bond $K_{\alpha}$ brings the factor $-\beta$ and enforces
the substitution $r' = r\sqrt{2\pi\beta n}$ which manifests itself 
as the factor $1/(2\pi\beta n)$ for each field-circle 
integration $\sim \int r \rd r$.
Since there are $(N_s-1)$ independent field-circle integrations in 
(\ref{26}) we conclude that
\begin{subequations}
\begin{equation} \label{27a}
{D_s(n) \over V} = n \beta^{L_s-N_s+1} d_s
\end{equation}
where $d_s$ is an intensive quantity, the number
\begin{equation} \label{27b}
d_s = {D_s(n=1,\beta=1) \over V}
\end{equation}
\end{subequations}
The first nonzero diagram from the sketch (\ref{11}) is $D_2$.
It contributes to the $\beta^3$ order, with
\begin{eqnarray} \label{28}
d_2 & = & {1\over 2! 4!} \int {\rd^2 r \over 2\pi} K_0^4({\vek r}) 
\nonumber \\
& = & {1\over 2! 4!}~ {7\over 8} ~ \zeta(3)
\end{eqnarray}
where $\zeta$ is the Riemann's zeta function (see Appendix).
In the next $\beta^4$ order only diagram $D_6$ survives, and
\begin{eqnarray} \label{29}
d_6 & = & {1\over 3! (2!)^3} \int {\rd^2 r_1 \over 2\pi}{\rd^2 r_2 \over 2\pi}
K_0^2({\vek r}_1) K_0^2({\vek r}_2) K_0^2({\vek r}_1-{\vek r}_2) \nonumber \\
& = & {1\over 3! (2!)^3}~ {3\over 16} ~ \zeta(3) 
\end{eqnarray}
(see Appendix), etc. 

\noindent The above paragraphs (i)--(iii) are summarized by formula
\begin{equation} \label{30}
{\Delta(n)\over V} = {\beta \over 4} (n \ln n - n ) + {\beta n \over 2}
\left[ C + {1\over 2} \ln \left( {\pi \beta \over 2} \right) \right]
+ n \sum_{s=1}^{\infty} d_s \beta^{L_s-N_s+1} 
\end{equation}
Here, $\{ d_s \}$ are the numbers yielded by the topology of diagram $D_s$,
nonzero only if the coordinations of all vertices are even numbers $\ge 4$.

In order to evaluate $\ln (n_{\sigma}/z_{\sigma})$ using (\ref{5}),
we first recall the well-known formula
\begin{equation} \label{31}
{\partial \Delta(n) \over \partial n} = \sum_{\sigma} \int \rd {\vek r}
{\delta \Delta[n] \over \delta n({\vek r},\sigma)} {\partial n({\vek r},\sigma)
\over \partial n} 
\end{equation}
valid for an arbitrary functional $\Delta[n]$ with $n({\vek r},\sigma)$
substituted by some function of $n$.
In the considered case $n({\vek r},+) = n({\vek r},-) = n/2$, relation
(\ref{31}) takes the form
\begin{equation} \label{32}
{\partial \Delta(n) \over \partial n} = {V\over 2} \left( {\delta \Delta[n] 
\over \delta n({\vek r},+)}\big\vert_{n({\vekexp r},\sigma)=n/2}+
{\delta \Delta[n] \over \delta n({\vek r},-)}\big\vert_{n({\vekexp r},\sigma)
=n/2} \right)
\end{equation}
where the reference ${\vek r}$-independence of the functional derivatives
was assumed.
For diagrams in $\Delta[n]$, the direct link between the $+$ and $-$ states
of the root point is realized through the state $\{ \sigma_i \to -
\sigma_i \}$ transformation at field vertices.
The diagrams are invariant with respect to this transformation, so that
it holds
\begin{equation} \label{33}
{\delta \Delta[n] \over \delta n({\vek r},+)}
\big\vert_{n({\vekexp r},\sigma)=n/2}= 
{\delta \Delta[n] \over \delta n({\vek r},-)}
\big\vert_{n({\vekexp r},\sigma)=n/2}=
{\partial \Delta(n)/V \over \partial n}
\end{equation}
The consequent equalities $\ln(n_+/z) = \ln(n_-/z) = 
\partial [\Delta(n)/V] / \partial n,$ with $\Delta(n)/V$ given by (\ref{30}),
lead after simple algebra to the final result
\begin{subequations} 
\begin{eqnarray} \label{34}
{n^{1-\beta/4} \over z} & = & 2 \beta^{\beta/4} \exp \left\{ 
\left[ 2C + \ln \left( {\pi \over 2} \right) \right] {\beta \over 4} 
+ \sum_{s=1}^{\infty} d_s \beta^{L_s-N_s+1} \right\} \label{34a} \\
& = & 2 \beta^{\beta/4} \exp \left\{ 
\left[ 2C + \ln \left( {\pi \over 2} \right) \right] {\beta \over 4}
+ {7\over 6} \zeta(3) \left( {\beta \over 4} \right)^3 + 
\zeta(3) \left( {\beta \over 4} \right)^4 + O(\beta^5) \right\} \label{34b}
\end{eqnarray}
\end{subequations}
The exact solution at the collapse border $\beta =2$ \cite{Cornu}
confirms the expected divergence $\sqrt{n}/z \to \infty$, and therefore the 
radius of convergence of the series 
$\sum_{s=1}^{\infty} \beta^{L_s-N_s+1} d_s$ corresponds to $0\le \beta
\le 2$.
In the next section, from the equivalence of the 2D TCP with the sine-Gordon 
model, the explicit form of the rhs of (\ref{34}) is found.

The internal consistency of the formalism presented in this part is
confirmed by the correct reproduction of the equation of state
for the 2D TCP \cite{Hauge},
\begin{equation} \label{35}
\beta P = n \left( 1- {\beta \over 4} \right)
\end{equation}
where $P$ is the pressure.
Writing down $\ln \Xi = \beta P V$ in Eq.(\ref{3}) and combining with
Eq.(\ref{4}), one obtains
\begin{equation} \label{36}
\beta P = {\Delta \over V} + n - n \ln \left( {n\over 2z} \right)
\end{equation}
The consequent substitution of (\ref{30}) and (\ref{34}) immediately
leads to the exact Eq.(\ref{35}).

\section{Mapping onto the sine-Gordon model}
It is well known that the 2D TCP is equivalent to the 2D Euclidean
(classical) sin-Gordon model \cite{Edwards}--\cite{Minnhagen}.
In particular, the grand partition function (\ref{1}) can be turned into
\begin{subequations} \label{37}
\begin{equation} \label{37a}
\Xi = { \int {\cal D} \phi ~ \exp \left( \int \rd^2 r ~ L \right) \over
\int {\cal D} \phi ~ \exp \left[ - \int \rd^2 r~ {1\over 2} \left( \nabla
\phi \right)^2 \right]}
\end{equation}
where
\begin{equation} \label{37b}
L =  - {1\over 2} \left( \nabla \phi \right)^2
+ 2 z \cos \left( \sqrt{2\pi \beta} ~ \phi \right) 
\end{equation}
\end{subequations}
The rescaling of the real field $\phi \to \phi/\sqrt{8\pi}$ transforms
$L$ into the form
\begin{subequations} \label{38}
\begin{eqnarray}
L & = & - {1\over 16 \pi} \left( \nabla \phi \right)^2
+ 2 z \cos \left( {\bar \beta} \phi \right) \label{38a} \\
{\bar \beta} & = & \sqrt{{\beta \over 4}} \label{38b}
\end{eqnarray}
\end{subequations}
which is more convenient for our purposes.
If we identify one of the two spatial coordinates with an imaginary time,
say $y = {\rm i} t$, $L$ (\ref{38a}) can be viewed as Lagrangian in the
$(1+1)$-dimensional Minkowski space.
Defining the canonical momentum conjugate to the field $\phi$,
\begin{equation} \label{39}
\Pi(x) = {\partial L \over \partial \dot{\phi}} = {1\over 8\pi} \dot{\phi} 
\end{equation}
the Hamiltonian density of the corresponding quantum $(1+1)$-dimensional
sine-Gordon model reads
\begin{equation} \label{40}
H = {1\over 16\pi} \left[ \Pi^2 + \left( \nabla \phi \right)^2 \right]
- 2 z \cos \left( {\bar \beta} \phi \right) 
\end{equation}
The thermodynamic properties of the 2D TCP at finite temperature may be 
therefore obtained from the ground-state properties of the corresponding
quantum sine-Gordon model, and vice versa.

The 2D Euclidean sine-Gordon theory (\ref{38}) has been intensively studied 
\cite{Faddeev}--\cite{Lukyanov}.
It can be regarded as the Gaussian Conformal FT perturbed by the relevant
operator $2\cos ({\bar \beta} \phi)$.
As concerns the renormalization of parameters ${\bar \beta}$ and $z$,
only $z$ renormalizes (multiplicatively).
Under the proposed normalization of the field $\cos ({\bar \beta}\phi)$
in terms of the short-distance limit of the two-point correlation function
\begin{equation} \label{41}
\langle \cos({\bar \beta} \phi)({\vek x}) \cos({\bar \beta} \phi)({\vek y}) 
\rangle \to {1\over 2} \vert {\vek x} - {\vek y} 
\vert^{-4{\bar \beta}^2} \quad {\rm as} \quad \vert {\vek x} - {\vek y} \vert
\to 0  
\end{equation}
the field $\cos({\bar \beta}\phi)$ has the dimension 
$[{\rm length}]^{-2{\bar\beta}^2}$, so that the dimension of parameter $z$
is $[{\rm length}]^{2{\bar\beta}^2-2}$, in full agreement with 
(\ref{34}) $z \sim n^{1-\beta/4}$ (the dimension of the density $n$ is
$[{\rm length}]^{-2}$ and ${\bar \beta}^2 = \beta/4$).
The discrete symmetry of the theory (\ref{38}) 
$\phi \to \phi + 2\pi n/{\bar \beta}$ ($n$ integer) is broken in
the region $0<{\bar \beta}<1$; one has to consider one of infinitely
many ground states $\{ \vert 0_n \rangle \}$ characterized by
$\langle \phi \rangle_n = 2\pi n/{\bar \beta}$, say that with $n=0$.
The underlying sine-Gordon model is integrable along the standard lines
of the Bethe ansatz technique \cite{Faddeev}, \cite{AZamolodchikov}.
The spectrum of particles involves solitons $S$, antisolitons ${\bar S}$
and soliton-antisoliton bound states $\{ B_j; j = 1, 2, \ldots < 1/\xi \}$,
\begin{equation} \label{42}
\xi = {{\bar \beta}^2 \over 1- {\bar \beta}^2} = {\beta \over 4-\beta}
\end{equation}
with masses
\begin{equation} \label{43}
m_j = 2 M \sin\left({\pi j \xi \over 2}\right)
\end{equation}
where $M$ is the soliton mass.
The dimensionless specific grand potential $\omega$, defined by
\begin{equation} \label{44}
\omega = - {1\over V} ~ \ln \Xi
\end{equation}
[and identical with the ground state energy of the quantum $(1+1)$-dimensional
sine-Gordon model] was found in Ref. \cite{Destri} (Eq. 2.67):
\begin{equation} \label{45}
\omega = - {m_1^2 \over 8 \sin(\pi \xi)}
\end{equation}
Mass $m_1$, associated with the lightest $B_1$-bound state, is given by
relation (\ref{43}), so that
$$ \omega = - {M^2 \over 4} \tan \left( {\pi \xi \over 2} \right) \eqno(45')$$
Within the framework of the normalization (\ref{41}), the parameter $z$
is related to the soliton mass $M$ as follows 
(see \cite{AlZamolodchikov}, Eqs. 2.12 and 4.1)
\begin{equation} \label{46}
z = {\Gamma\left( {\bar \beta}^2 \right) \over \pi \Gamma\left( 1- 
{\bar \beta}^2 \right) } \left[ M {\sqrt{\pi} \Gamma\left( {1\over 2}
+ {\xi \over 2} \right) \over 2 \Gamma\left( {\xi\over 2} \right)}
\right]^{2-2{\bar \beta}^2}
\end{equation}
where $\Gamma$ stands for the gamma function.
Notice a different parameter notation in Refs. \cite{Destri}, 
\cite{AlZamolodchikov}.

Returning to the $\beta$-parameter [see formulae (\ref{38b}), (\ref{42})]
and applying the transformation relation between the Gamma functions 
\cite{Gradshteyn}
\begin{equation} \label{47}
\Gamma(x) = {1\over x} \Gamma(1+x)
\end{equation}
the elimination of $M$ from Eq.(\ref{46})
implies the explicit $z$-dependence of $\omega$,
\begin{equation} \label{48}
-\omega(z,\beta) = \left( 1 - {\beta \over 4} \right) z^{{1\over 1-\beta/4}}
\left[ 2 \left( {\pi \beta \over 8} \right)^{\beta/4}
{\Gamma \left( 1-{\beta\over 4} \right) \over \Gamma \left( 1+{\beta\over 4} 
\right)} \right]^{{1\over 1-\beta/4}}
{\tan \left[ {\pi \beta \over 2(4-\beta)} \right] \over
{\pi \beta \over 2(4-\beta)}} { \Gamma^2 \left[ 1 + {\beta \over 
2(4-\beta)} \right] \over {1\over \pi} \Gamma^2 \left[ {1\over 2} + 
{\beta \over 2(4-\beta)} \right]} 
\end{equation}
Inserting then $(-\omega)$ into the generating relation for 
the particle density
\begin{equation} \label{49}
n = z {\partial (-\omega) \over \partial z}
\end{equation}
one finally obtains
\begin{equation} \label{50}
{n^{1-\beta/4}\over z} = 
2 \left( {\pi \beta \over 8} \right)^{\beta/4} {\Gamma \left( 1-{\beta\over 4}
\right) \over \Gamma \left( 1+{\beta\over 4} \right)} 
\left\{ {\tan \left[ {\pi \beta \over 2(4-\beta)} \right] \over
{\pi \beta \over 2(4-\beta)}} { \Gamma^2 \left[ 1 + {\beta \over 
2(4-\beta)} \right] \over {1\over \pi} \Gamma^2 \left[ {1\over 2} + 
{\beta \over 2(4-\beta)} \right]} \right\}^{1-\beta/4}
\end{equation}
The rhs of (\ref{50}) predicts the stability against collapse for 
$\beta < 2$ and the divergency of $n$ at $\beta \to 2$, as it should be.
Its expansion around $\beta = 0$ up to the $\beta^4$ term, accomplished
e.g. by using the symbolic computer language {\it Mathematica}, 
turns out to be identical to our previous result (\ref{34b}).
This is the strong evidence that formula (\ref{50}) reflects adequately
the diagrammatic series $\sum_{s=1}^{\infty} \beta^{L_s-N_s+1} d_s$ in
(\ref{34a}), and represents the exact density-fugacity relationship
for the 2D TCP.

We would like to stress that there were attempts in the past to
get the thermodynamics of the 2D TCP from the mapping onto the
sine-Gordon theory, but they failed because of an inadequate transfer
of the information.
We mention one earlier work \cite{Saglio}, the result of which was proved 
to be wrong in report \cite{Olaussen}.
The crucial point here is that the dependence of the parameter $z$ on
the soliton mass $M$, relation (\ref{46}), was determined within the
field normalization (\ref{41}), the last being consistent with the
formalism of the renormalized Mayer expansion (Sections 3 and 4).

\section{Thermodynamics of the 2D TCP}
The simplest way to get the thermodynamics of the 2D TCP is to pass
from the grandcanonical to the canonical ensemble via the Legendre
transformation
\begin{equation} \label{51}
F(T,N) = \Omega + \mu N
\end{equation}
where
\begin{subequations} 
\begin{eqnarray} 
\Omega & = & k_B T ~ \omega(\beta,n) V  \label{52a} \\
- \omega(\beta,n) & = & \left( 1- {\beta \over 4} \right) n \label{52b}
\end{eqnarray}
\end{subequations}
and 
\begin{equation} \label{53}
\mu(\beta,n) = k_B T ~ \ln z(\beta,n)
\end{equation}
The dimensionless specific free energy $f$, defined by
$f = F/(Nk_BT)$, is then written as
\begin{eqnarray} \label{54}
f(\beta,n) & = & - \left( 1-{\beta \over 4} \right) +
\left( 1-{\beta \over 4} \right) \ln n - \ln \left[ 2 \left( {\pi \beta 
\over 8} \right)^{\beta/4} {\Gamma \left( 1-{\beta\over 4} \right) \over 
\Gamma \left( 1+{\beta\over 4} \right)}\right]  \nonumber \\
& & - \left( 1-{\beta\over 4} \right) \ln \left\{ {\tan \left[ {\pi \beta 
\over 2(4-\beta)} \right] \over {\pi \beta \over 2(4-\beta)}} { \Gamma^2 
\left[ 1 + {\beta \over 2(4-\beta)} \right] \over {1\over \pi} \Gamma^2 
\left[ {1\over 2} + {\beta \over 2(4-\beta)} \right]} \right\} 
\end{eqnarray}
According to the elementary thermodynamics, the internal energy per particle,
$e = \langle E \rangle /N$, and the specific heat (at constant volume) per
particle, $c_V = C_V/N$, are given by
\begin{subequations}
\begin{eqnarray}
e & = & {\partial \over \partial \beta} f(\beta,n) \label{55a} \\
{c_V\over k_B} & = & - \beta^2 {\partial^2 \over \partial \beta^2}
f(\beta,n) \label{55b}
\end{eqnarray}
\end{subequations}
For the specific heat, one gets explicitly
\begin{eqnarray} \label{56}
{c_V\over k_B} & = & {\beta \over 4} + {4\over 4-\beta} + {\beta^2 \over 16}
\left[ \psi^{(1)}\left( 1-{\beta\over 4} \right) - \psi^{(1)}\left( 1+
{\beta\over 4} \right)\right] \nonumber \\
& & -2 {\beta^2 \over (4-\beta)^3} \left[ \psi^{(1)}\left( {2\over 4-\beta}
\right) - \psi^{(1)}\left( {8-\beta \over 8-2\beta}\right)\right] -
{4\pi^2 \beta^2 \over (4-\beta)^3} {\cos\left( {\pi \beta \over 4-\beta}\right)
\over \sin^2\left( {\pi \beta \over 4-\beta}\right)}
\end{eqnarray}
The series representation of the psi function and of its derivatives
is presented in Appendix, formula (A7). 
It is seen that the density-independent $c_V/k_B$ has a regular 
high-temperature expansion in powers of $\beta$:
\begin{equation} \label{57}
{c_V\over k_B} = {\beta \over 4} + {7\over 64} \zeta(3) \beta^3
+ {3\over 64} \zeta(3) \beta^4 + O(\beta^5)
\end{equation}
The expansion of $c_V/k_B$ around the collapse $\beta =2$ point 
results in the Laurent series
\begin{equation} \label{58}
{c_V \over k_B} = {2\over (2-\beta)^2} - {3\over (2-\beta)} + {3\over 2}
+ {1\over 4} \left[ 17 \zeta(3) -1 \right] (2-\beta) + O[(2-\beta)^2]
\end{equation}
The leading term coincides exactly with the conjecture of Hauge and Hemmer
\cite{Hauge} [see their Eq.(34)] based on an independent-pair approximation.

In conclusion, it was shown that the renormalized reformulation of the
Mayer expansion in statistical mechanics of the 2D TCP produces a direct
link to techniques in the related QFT.
The obtained results might represent a step towards the
complete integrability of the 2D TCP on higher correlation levels.
A further exploration of the intimate relationship to recent studies
of the sine-Gordon model is needed.

\section*{Acknowledgment}
We are indebted to Bernard Jancovici and Fran\c{c}oise Cornu for
their interest in our work and clarifying discussion about a negligible 
effect of non-neutral configurations in plasma.
We are grateful to Bernard Jancovici also for the careful reading of 
the manuscript and very useful comments, for providing us with Ref.
\cite{Olaussen} and with his previous correspondence concerning the topic. 
This work was supported by grant VEGA no 2/7174/20.

\newpage

\renewcommand{\theequation}{A\arabic{equation}}
\setcounter{equation}{0}

\section*{Appendix}
We aim at proving the relations:
\begin{subequations}
\begin{eqnarray} \label{A1}
\int {\rd {\vek r} \over 2\pi} K_0^4({\vek r}) & = & {7\over 8} ~ \zeta(3) 
\label{A1a} \\
\int {\rd {\vek r}_1 \over 2\pi}{\rd {\vek r}_2 \over 2\pi}
K_0^2({\vek r}_1) K_0^2({\vek r}_2) K_0^2({\vek r}_1-{\vek r}_2) & = & 
{3\over 16} ~ \zeta(3) \label{A1b}
\end{eqnarray}
\end{subequations}
\medskip
The Fourier component of $K_0^2({\vek r})$, denoted as $G({\vek k})$,
can be expressed in 2D as follows
\begin{eqnarray} \label{A2}
G({\vek k}) & = & \int {\rd^2 r \over 2\pi} \exp \left( {\rm i} {\vek k}
\cdot {\vek r} \right) K_0^2({\vek r}) \nonumber \\
& = & \int_0^{\infty} \rd r ~ r J_0(kr) K_0^2(r) \nonumber \\
& = & {\ln \left[ {k\over 2} + \sqrt{1+\left({k\over 2}\right)^2}\right]
\over k ~ \sqrt{1+\left({k\over 2}\right)^2}}
\end{eqnarray}
where $J_0$ is the ordinary Bessel function.
Simple algebra yields
\begin{subequations}
\begin{eqnarray} \label{A3}
\int {\rd {\vek r} \over 2\pi} K_0^2({\vek r}) K_0^2({\vek r}) & = & 
\int_0^{\infty} \rd k ~ k G^2(k) \label{A3a} \\
\int {\rd {\vek r}_1 \over 2\pi}{\rd {\vek r}_2 \over 2\pi}
K_0^2({\vek r}_1) K_0^2({\vek r}_2) K_0^2({\vek r}_1-{\vek r}_2) & = & 
\int_0^{\infty} \rd k ~ k G^3(k) \label{A3b}
\end{eqnarray}
\end{subequations}
The primitive functions of the functions $k G^2(k)$ and $k G^3(k)$
are available explicitly e.g. with the aid of {\it Mathematica}. 
They converge to 0 at $k\to\infty$, and their values in the $k\to 0$
limit imply
\begin{subequations}
\begin{eqnarray} \label{A4}
\int_0^{\infty} \rd k ~ k G^2(k) & = & {\zeta(3) \over 2} -
2 \left\{ {\rm PolyLog}[3,{\rm i}]+{\rm PolyLog}[3,-{\rm i}] \right\} 
\label{A4a}\\
\int_0^{\infty} \rd k ~ k G^3(k) & = & {3\over 4} ~ \zeta(3)
+ 3 \left\{ {\rm PolyLog}[3,{\rm i}]+{\rm PolyLog}[3,-{\rm i}] \right\} 
\label{A4b}
\end{eqnarray}
\end{subequations}
where
\begin{equation} \label{A5}
{\rm PolyLog} [n,x] = \sum_{j=1}^{\infty} {x^j \over j^n}
\end{equation}
is the $n$th polylogarithm function of $x$.
It is straigthforward to show that
\begin{equation} \label{A6}
{\rm PolyLog}[3,{\rm i}]+{\rm PolyLog}[3,-{\rm i}] = 
{1\over 32} \left[ \zeta(3) - \sum_{j=1}^{\infty} {1\over \left( 
j-{1\over 2} \right)^3} \right]
\end{equation}
The $n$-th derivative of the psi function is given by
\begin{equation} \label{A7}
\psi^{(n)}(x) = (-1)^{n+1} n! \sum_{j=0}^{\infty} {1\over (x+j)^{n+1}}
\end{equation}
Deriving twice the functional relation \cite{Gradshteyn}
\begin{equation} \label{A8}
\psi(2 x) = {1\over 2} \left[ \psi(x) + \psi\left( x+{1\over 2}\right) \right]
+ \ln 2 
\end{equation}
at $x=1/2$, we have $\psi^{(2)}(1/2) = 7 \psi^{(2)}(1)$. Thus,
\begin{equation} \label{A9}
\sum_{j=1}^{\infty} {1\over \left( j-{1\over 2} \right)^3} =
- {1\over 2} \psi^{(2)}\left( {1\over 2}\right) = - {7\over 2}
\psi^{(2)}(1) = 7 \zeta(3) 
\end{equation}
where we have used the definition of the Riemann's zeta function
$\zeta(3) = - \psi^{(2)}(1)/2$.
The insertion of the consequent equality
\begin{equation} \label{A10}
{\rm PolyLog}[3,{\rm i}]+{\rm PolyLog}[3,-{\rm i}] = - {3\over 16}
~ \zeta(3)
\end{equation}
into (A4), and the consideration of Eqs.(A3) proves (A1).

\end{document}